\documentclass[preprint,authoryear,12pt]{elsarticle}
\usepackage{amssymb}
\newtheorem{thm}{Theorem}
\newtheorem{lem}{Lemma}
\newtheorem{pr}{Proposition}
\newdefinition{rmk}{Remark}
\newdefinition{ex}{Example}
\newproof{pf}{Proof}
\newproof{pot1}{Proof of Theorem \ref{thm1}}
\newproof{pot2}{Proof of Theorem \ref{thm2}}

\begin{document}

\begin{center}
{\Large\bf Stochastic Optimal Growth Model with Risk Sensitive Preferences}
\end{center}

\begin{center}{\bf \large Nicole B\"auerle $^{a}$, Anna Ja\'skiewicz$^{b}$ } \end{center}
\noindent$^{a}$Department of Mathematics, Karlsruhe Institute of Technology, Karlsruhe, Germany
{\footnotesize {\it email: nicole.baeuerle@kit.edu}}\\
\noindent$^{b}$Faculty of Mathematics, Wroc{\l}aw University of Technology,  Wroc{\l}aw, Poland
{\footnotesize {\it email: anna.jaskiewicz@pwr.edu.pl}}\\

\noindent {\bf Abstract.} This paper studies a one-sector optimal growth model with i.i.d. productivity shocks that are allowed
to be unbounded. The utility function  is assumed to be non-negative and unbounded from above. The novel feature in our framework
is that the agent has risk sensitive preferences in sense of \cite{hs}. Under mild assumptions imposed on the productivity and utility
functions we prove that the maximal discounted non-expected utility in the infinite time horizon satisfies the optimality equation and the
agent possesses a stationary optimal policy.  A new point used in our analysis is an inequality for the so-called associated random variables.
We also establish the Euler equation that incorporates the solution to the optimality equation.\\


\section{Introduction}

This paper deals with one-sector stochastic optimal growth model with possibly unbounded shocks and  non-negative utilities that are
allowed to be unbounded from above. Unbounded returns are very common in economic models,
see \cite{as, boyd,d1,lvm} for the deterministic problems, and \cite{d2,jn,kami}
for stochastic problems. Most of the aforementioned works apply the weighted supremum norm approach introduced by \cite{w}.
The other group of papers makes use of the idea presented by \cite{zec} within deterministic framework.
Their method rests upon a local contraction and utilises one-sided majorant functions. The extensions of these results
to stochastic dynamic programming are reported in \cite{jn_jmaa, rv,matnow}.

The novelty in our model relies on the fact that  the agent has risk sensitive preferences of the form
\begin{equation}
\label{hs_pr}
V_t=u(a_t)-\frac\beta\gamma \ln E_t[-\gamma V_{t+1}],
\end{equation}
where $\gamma>0$ is a risk sensitive coefficient, $\beta\in[0,1)$ is a time discount factor, $a_t$ is consumption at time $t,$
$u$ is a felicity function and $V_t$ is the lifetime utility from period $t$ onward. Here, $E_t$ stands for the expectation operator
with respect to period $t$ information. The parameter $\gamma$ affects  consumer's attitude towards risk in future utility.
The form of  preferences in (\ref{hs_pr}) is due to \cite{hs}, who used them to deal with a linear quadratic Gaussian control model.
 The preferences  defined in  (\ref{hs_pr}) has several advantages. First of all,  they are not
 time-additive in future utility. Time-additivity, however, requires an agent
to be risk neutral in future utility.
Risk sensitive preferences, on the other hand,  allow the agent to be risk
averse in future utility in addition to being risk averse in future consumption.
This fact results in partial separation between risk aversion and elasticity
of intertemporal substitution.\footnote{Consult \cite{ez}, \cite{tal}, \cite{miao} for more discussion on this topic.
The reader is also referred to Chapters 1 and 3 in \cite{bb}that  constitute
a strong motivation for the study of non-time-additive objective functions.}
Moreover, as argued by \cite{hst} risk sensitive preferences are also attractive, because they can  be used
to model preferences for robustness.
 It is worth emphasising that risk sensitive preferences of form (\ref{hs_pr}) have found several applications, for instance,
in  the problems dealing with Pareto optimal allocations
 (see \cite{a}) or small noise expansions (see \cite{ahs}).

Our main results are two-fold. First
we establish the optimality equation for the non-expected utility in the infinite time horizon,
when the agent has risk sensitive preferences (\ref{hs_pr}).
The proof as in the standard expected utility case is based on the Banach contraction principle, see \cite{slp}.
However, in oder to show that the dynamic programming operator maps a space of certain functions into itself,
we have to confine our consideration to concave, non-decreasing and non-negative functions
that are bounded in the weighted supremum norm.  A novel feature in this analysis is an application of some inequality
for the so-called associated random variables. This inequality also plays a crucial role in proving that the a fixed point
of dynamic programming operator is indeed the value function. Moreover,  it naturally fits into our model, in which
the production and utility functions satisfy some mild conditions such as
monotonicity and concavity. Here, we would like to emphasise that similarly as in \cite{kami},
we do not assume the Inada conditions for the production function at zero and infinity.
Secondly, we establish the Euler equation assuming that the production and utility functions are continuously differentiable.
As an by-product, we obtain the existence of a distribution function for the income process governed by the optimal policy.
This result owes much previous works by  \cite{ns_jet}, \cite{st}  that link  the Euler equation with
the Foster- Lyapunov functions.

The paper is organised as follows. Section 2 describes the model,
risk sensitive preferences of the agent and provides essential assumptions.
In Section 3, we use the dynamic programming approach to show
that the agent has an optimal stationary policy and the lifetime utility is
a solution to the optimality (Bellman) equation. Section 4 establishes the Euler equation.
Section 5 makes use of the Euler equation to define a Foster-Lyapunov function, which is applied
to the proof of stability of the optimal program. Finally, Section 6 gives  two examples that illustrate our theory.

\section{The  Model}

This section contains a formulation of the stochastic optimal growth model with the payoff criterion that in a
particular case  reduces to the one studied by \cite{bm}.
The symbols $\mathbb{R}_+$ and $\mathbb{R}_{++}$ denote the non-negative and positive real numbers,
respectively. Let $\mathbb{N}$ be the set of positive integers.
The process evolves as follows. At time $t\in\mathbb{N}$ the agent has an income $x_t,$
which is divided between consumption $a_t$ and investment (saving) $y_t.$ From consumption $a_t$
the agent receives utility $u(a_t)$ (which is independent of $x_t$). Investment is used  for
production with input $y_t$ yielding output $x_{t+1}=f(y_t,\xi_{t}),$ where $(\xi_t)_{t\in\mathbb{N}}$
is a sequence of i.i.d. shocks with distribution $\nu\in \Pr(\mathbb{R}_+)$ and $f:\mathbb{R}_+
\times \mathbb{R}_+\mapsto \mathbb{R}_+$ is a production function. It is assumed that
$x_1\in \mathbb{R}_{+}$ is non-random.

We make the following assumptions.

\begin{enumerate}
  \item[(U1)] The function $u:\mathbb{R}_+\mapsto \mathbb{R}_+$ is continuous at zero, strictly concave, increasing and
  $u(0)=0.$
  \item[(U2)] There exist a constant $d>0$ and
  a continuous non-decreasing function $w: {\mathbb R}_+ \mapsto [1,\infty)$ such that
  $$u(x)\le dw(x),\quad \mbox{for all }x\in {\mathbb R}_+$$
  and (\ref{F2})  holds.
  \item[(F1)] For every $z\ge 0$ the function $f(\cdot,z):\mathbb{R}_+ \mapsto \mathbb{R}_+$ is
  continuous, concave, non-decreasing and for every $y\ge 0$ the function
  $f(y,\cdot):\mathbb{R}_+ \mapsto \mathbb{R}_+$ is Borel measurable.
  \item[(F2)] There exists a constant $\alpha\in (0, 1/\beta)$ such that
\begin{equation}
\label{F2}
\sup_{y\in [0,x]}\int_{\mathbb{R}_+} w(f(y,z))\nu(dz)\le \alpha w(x) ,
  \quad \mbox{for all }x\in {\mathbb R}_+.
  \end{equation}
 \end{enumerate}

Put
$$D:=\{(x,y):\; x\in\mathbb{R}_+,\ y\in[0,x]\}.$$
For any $t\in\mathbb{N},$  by $H_t$ we denote the set of all sequences
$$h_t=\left\{\begin{array}{l@{\quad \quad}l}
x_1,& \mbox{ for } t=1,\\
 (x_1, y_1,x_2,\ldots, x_t), & \mbox{ for } t\ge2,
 \end{array}\right.$$
where $(x_k,y_k)\in D$  for all  $k\in \mathbb{R}_+.$
Hence, $H_t$ is the set of all feasible  histories of the income-investment process up to date $t$.
An  {\it investment policy} $\pi$ is a sequence $(\pi_t)_{t\in \mathbb{N}},$ where $\pi_t$ is a measurable mapping
which associates any admissible history $h_t$ with an action $y_t \in [0,x_t]$.
By $\Pi$ we denote the set of {\it all investment policies}.  We restrict our attention to
non-randomised policies, which are enough to study the discounted optimal growth models.
A  formal definition of a general policy can be found in \cite{slp}.
Let $\Phi$ be the set of all Borel measurable functions such that
$\phi(x)\in [0,x]$ for every $x\in \mathbb{R}_+.$
A {\it stationary (investment) policy} is a constant sequence $\pi$ with $\pi_t =\phi$ for every $t\in \mathbb{N}.$
We shall identify a stationary policy with the member of the sequence, i.e., with the mapping $\phi.$ By
$\Phi$ we also  denote the set of all stationary investment policies.

In this paper, we shall consider the {\it non-expected utility} of the agent in the infinite time horizon
derived with the aid of the so-called entropic risk measure.  In order to define this measure let us denote by
 $(\Omega,{\cal F}, P)$ a probability space and let $X$
be a non-negative random payoff defined on $(\Omega,{\cal F}, P)$.
The entropic risk measure of $X$ is
$$\rho(X)= -\frac 1\gamma\ln\int_{\Omega} e^{-\gamma X(\omega)}P(d\omega),$$
where $\gamma>0$ is the {\it risk sensitive  coefficient.  }
Let $X,$ $Y$ be non-negative random variables on $(\Omega,{\cal F}, P)$.
The following properties of $\rho$ are important in the analysis (see p. 184 in  \cite{fs}):
\begin{enumerate}
  \item[(P1)] monotonicity, i.e., if $X\le Y$ $\Rightarrow$ $\rho(X)\le \rho(Y)$
  \item[(P2)] concavity, i.e., $\rho(\lambda X+(1-\lambda)Y)\ge \lambda\rho(X)+(1-\lambda)\rho(Y)$
  for any $\lambda\in[0,1].$
  \end{enumerate}
  However, this measure is not positive homogeneous, i.e., $\rho(\alpha X)\not=\alpha\rho(X)$ for $\alpha\in\mathbb{R}_{++},$ which does not make
  our analysis straightforward (see the proofs of Lemma 4 and Theorem 1).
Here, we wish only to mention that making use of the Taylor expansions for the exponential and logarithmic functions, we can approximate
$\rho(X)$  as follows
$$  \rho(X)\approx EX-\frac\gamma 2Var X,$$
if $\gamma>0$ is sufficiently close to $0.$
Therefore, if $X$ is a random payoff,
then  the agent who evaluates his expected payoff with the aid of the entropic risk measure,
thinks not only of
the expected value $EX$ of the random payoff $X$, but also
 of its variance. Further comments on the entropic risk measure can be found in \cite{bra, fs} and references cited therein.

Assume that  $k\in\mathbb{N}.$
We say that a function $v_k\in B_w(H_k),$ if $v_k: H_k\mapsto  \mathbb{R}_+$
is  Borel measurable and there exists a constant $d_{v_k}\ge 0$ such that $v_k(h_k)\le d_{v_k}w(x_k)$ for every   $h_k\in H_k.$
Here, $w$ is a function used in (U2) and (F2).
Let  $\pi=(\pi_k)_{k\in\mathbb{N}}\in\Pi$  be any policy.
For   $v_{k+1}\in B_w(H_{k+1})$ and  given $h_k\in H_k$  we put
\begin{equation}
\label{ece}
\rho_{\pi_k,h_k}(v_{k+1}):= -\frac 1\gamma\ln\int_{\mathbb{R}_+} e^{-\gamma v_{k+1}(h_k,\pi_k(h_k), f(\pi_k(h_k),z))}\nu(dz).
\end{equation}
Observe that  by Jensen's inequality and (F2)  we have
\begin{eqnarray}
\label{rho_ogr}
0&\le&
\rho_{\pi_k,h_k}(v_{k+1})
\le \int_{\mathbb{R}_+}v_{k+1}(h_k,\pi_k(h_k), f(\pi_k(h_k),z))\nu(dz)\\ \nonumber
&\le&
d_{v_{k+1}}\int_{\mathbb{R}_+}w(f(\pi_k(h_k),z))\nu(dz)\\ \nonumber
&\le&
d_{v_{k+1}}\sup_{y\in[0,x_k]}\int_{\mathbb{R}_+}w(f(y,z))\nu(dz)\le
d_{v_{k+1}} \alpha w(x_k)
\end{eqnarray}
for any $h_k\in H_k$ and  $k\in\mathbb{N}.$ Furthermore,  we define the operator
$L_{\pi_k}$ as follows
$$L_{\pi_k}v_{k+1}(h_k):=u(x_k-\pi_k(h_k))+\beta\rho_{\pi_k,h_k}(v_{k+1}),$$
where $\beta\in (0,1)$ is a subjective discount factor.  By property (P1), it follows that $L_{\pi_k}$ is monotone, i.e.,
$L_{\pi_k}v_{k+1}(h_k)\le L_{\pi_k}\hat{v}_{k+1}(h_k)$ for $h_k\in H_k$ and $v_{k+1}\le \hat{v}_{k+1}.$
Moreover, by (\ref{rho_ogr}), (U2) and (F2) we get that
\begin{equation}
\label{L_ogr}
0\le L_{\pi_k}v_{k+1}(h_k)\le (d+\alpha\beta d_{v_{k+1}})w(x_k)
\end{equation}
for every $h_k\in H_k$ with $k\in\mathbb{N}.$

We follow the approach of \cite{hs} and model the preferences of the consumer recursively.
For any initial income $x_1=x$ and  $T\in {\mathbb N}$ we define $T$-stage total discounted utility
\begin{equation}\label{J_T}
J_T(x,\pi):=(L_{\pi_1}\circ \ldots \circ L_{\pi_T} ){\bf 0}(x),
\end{equation}
where  ${\bf 0}$ is a function such that  ${\bf 0}(h_k)\equiv 0$ for every  $h_k\in H_k$ and $k\in\mathbb{N}.$
For instance, if $T=2$ definition (\ref{J_T}) is read as follows
\begin{eqnarray*}
J_2(x,\pi)&=&
(L_{\pi_1}\circ L_{\pi_2}){\bf 0}(x)=L_{\pi_1}( L_{\pi_2}{\bf 0})(x)\\
&=& u(x-\pi_1(x))-\frac\beta\gamma\ln\int_{\mathbb{R}_+} e^{-\gamma  L_{\pi_2}{\bf 0}(x,\pi_1(x),f(\pi_1(x),z))}\nu(dz)\\
&=&u(x-\pi_1(x))-\frac\beta\gamma\ln\int_{\mathbb{R}_+} e^{-\gamma u(f(\pi_1(x),z)- \pi_2(x,\pi_1(x),f(\pi_1(x),z)))}\nu(dz).\\
\end{eqnarray*}
Observe that  from (U1) and (P1), it follows that the sequence $(J_T(x,\pi))_{T\in\mathbb{N}}$ is non-decreasing and
bounded from below by $0$   for every $x\in{\mathbb R}_+$ and $\pi\in\Pi.$
Moreover,
$$
J_T(x,\pi)\le \frac {dw(x)}{1-\alpha\beta}\quad\mbox{ for } \quad x\in\mathbb{R}_+,\,\, \pi\in\Pi,\,\, T\in\mathbb{N}.
$$
Indeed, note first that by (U2) we have
\begin{equation}\label{111}
L_{\pi_T}{\bf 0}(h_T)=u(x_T-\pi_T(h_T))\le u(x_T)\le dw(x_T)\le \frac{dw(x_T)}{1-\alpha\beta},\quad h_T\in H_T.
\end{equation}
Now making use of (\ref{L_ogr}) with $k:=T-1,$ $v_T(h_T):=dw(x_T)/(1-\alpha\beta),$ (\ref{111})
and monotonicity of the operator  $L_{\pi_{T-1}}$
 we obtain
 $$ L_{\pi_{T-1}} (L_{\pi_T} {\bf 0})(h_{T-1})\le  L_{\pi_{T-1}} \left(\frac{dw}{1-\alpha\beta}\right)(h_{T-1})\le
dw(x_{T-1})+\alpha\beta\frac{dw(x_{T-1})}{1-\alpha\beta}=\frac{dw(x_{T-1})}{1-\alpha\beta}.$$
Continuing this procedure, we finally infer that
\begin{eqnarray*}
 J_{T}(x,\pi)&=&(L_{\pi_1}\circ \ldots \circ L_{\pi_T} ){\bf 0}(x)
\le (L_{\pi_1}\circ \ldots \circ L_{\pi_{T-2}} )\left(\frac{dw}{1-\alpha\beta}\right)(x)\le\ldots\\
&\le& d w(x)+ \frac{\alpha\beta  dw(x)}{1-\alpha \beta}= \frac{dw(x)}{1-\alpha \beta}.
\end{eqnarray*}
By the above discussion $ \lim_{T\to\infty} J_T(x,\pi)$ exists for every $x\in \mathbb{R}_+$ and $\pi\in\Pi.$\\

\noindent
{\it The problem statement.} For an initial income $x\in\mathbb{R}_+$ and policy $\pi\in\Pi$
we define the non-expected discounted utility in the infinite time horizon as follows
\begin{equation}
\label{J}
J(x,\pi):= \lim_{T\to\infty} J_T(x,\pi).
\end{equation}
The aim of the agent is to find an optimal value (the so-called value function) of
the non-expected discounted utility in the infinite time horizon and a policy  $\pi^*\in \Pi$ for which
$$J(x,\pi^*)=\sup_{\pi\in\Pi} J(x,\pi), \quad \mbox{for all } x\in{\mathbb R}_+.$$

\begin{rmk}
When the risk sensitive coefficient $\gamma\to 0^+,$ then the  non-expected utility  in (\ref{J})
tends to the von Neumann-Morgenstern  expected utility that was first studied by \cite{bm}
for a stochastic optimal growth model.
The greater $\gamma>0$   the more risk averse is the agent.

The entropic risk measure, which we used in our framework (see (\ref{ece}) is also  known as the exponential certainty equivalent.
It can be used to study the models with robust preferences. This is because,  this measure has a robust representation with
the relative entropy as a penalty function, see \cite{fs} or \cite{jn}.
Moreover, as noted by \cite{hs} the specification of such recursion provides a bridge between risk sensitive control theory (see \cite{wh})
 and a more general recursive utility specification used by \cite{ez}.\footnote{The reader is referred to Chapter 20 in  \cite{miao},
 where a detailed discussion, further references
on this topic  are provided.}
\end{rmk}

\section{The Bellman Equation}

In order to solve the aforementioned problem we shall use the dynamic programming approach.
We start from the definition of a class of functions among which we look for a solution to the optimality equation.

For a Borel measurable function $v: {\mathbb R}_+\mapsto {\mathbb R}$ define its $w$-norm as follows
$$\| v\|_w := \sup_{x\in\mathbb{R}_+}\frac{|v(x)|}{w(x)}.$$
Let $B_w$ be the set of all Borel measurable functions $v: {\mathbb R}_+\mapsto {\mathbb R}$ with the finite
$w$-norm. Then, $(B_w,\|\cdot\|_w)$ is a Banach space (see Proposition 7.2.1 in \cite{hl}).
Define
$$ {\cal B}:= \left\{v\in B_w: v \;\mbox{is continous,
concave, non-decreasing, non-negative} \right\}.$$
Note that $({\cal B},\|\cdot\|_w)$ is a complete metric space as a closed subset of the Banach space
$(B_w,\|\cdot\|_w).$

Now we are ready to state our first result.

\begin{thm} \label{thm1} Assume (U1)-(U2) and (F1)-(F2). Then, the following holds.
\begin{itemize}
  \item[(a)] There exist unique functions $V\in {\cal B}$ and $i^*\in \Phi$ such that
  \begin{eqnarray}
  \label{oe}
  V(x)&=&\sup_{y\in[0,x]}\left(u(x-y)-
  \frac\beta\gamma\ln\int_{{\mathbb R}_+}e^{-\gamma V(f(y,z))}\nu(dz)\right)\\
   \label{oe_max}
  &=& u(x-i^*(x))-
  \frac\beta\gamma\ln\int_{{\mathbb R}_+}e^{-\gamma V(f(i^*(x),z))}\nu(dz)
  \end{eqnarray}
  for all $x\in {\mathbb R}_+.$ Moreover, $V$ is strictly concave.
  \item[(b)]  The functions $x\mapsto i^*(x)$ and
  $x\mapsto c^*(x):=x-i^*(x)$ are continuous and non-decreasing.
   \item[(c)] $V(x)=\sup_{\pi\in \Pi} J(x,\pi)=J(x,i^*)$ for all $x\in {\mathbb R}_+$, i.e.\ there exists an optimal stationary policy $i^*$.
\end{itemize}
\end{thm}

Throughout this section we assume that (U1)-(U2) and (F1)-(F2) are satisfied.
We start with a result that we shall use in many places.

\begin{lem} \label{basic}
Let $v\in {\cal B}.$ Then, the function
$$y\mapsto \widehat{v}(y):=-\frac 1\gamma \ln\int_{{\mathbb R}_+}e^{-\gamma v(f(y,z))}\nu(dz)$$
is continuous, concave, non-decreasing and non-negative.
\end{lem}

\begin{pf}
By (F1) we have $0\le f(0,z)$ and $0\le v(0)\le v(f(0,z))$ for any $z\in{\mathbb R}_+.$ Hence, (P1)
yields that $0\le \widehat{v}.$
Furthermore, by (F1) the function $v(f(\cdot,z))$ is non-decreasing  for every $z\in\mathbb{R}_+.$
Thus, by (P1) the function $\widehat{v}$ is also non-decreasing.
Similarly, making use again of (F1) we find that  $v( f(\cdot,z))$ is continuous for every
$z\in\mathbb{R}_+.$ Hence, the dominated convergence theorem implies that $\widehat{v}$ is continuous.
Finally, we show the concavity of $\widehat{v}.$ Let $y=\lambda y'+ (1-\lambda) y'',$
where $\lambda\in(0,1).$ By (F1) it follows that
$$f(y,z)\ge \lambda f(y',z)+ (1-\lambda) f(y'',z),\quad z\in\mathbb{R}_+,$$
and by  the fact that $v$ is non-decreasing and concave we obtain
$$v(f(y,z))\ge v(\lambda f(y',z)+ (1-\lambda) f(y'',z))\ge
\lambda v(f(y',z)+ (1-\lambda) v(f(y'',z)),\quad z\in\mathbb{R}_+.$$
Now, properties (P1) and (P2) imply that
\begin{equation}
\label{oper_rho}
\widehat{v}(y)\ge -\frac 1\gamma
  \ln\int_{{\mathbb R}_+}e^{-\gamma (\lambda v(f(y',z)+ (1-\lambda) v(f(y'',z)))}\nu(dz)
  \ge
  \lambda \widehat{v}(y')+ (1-\lambda) \widehat{v}(y''),
\end{equation}
which finishes the proof.\hfill $\Box$
\end{pf}

\begin{lem} \label{incr}
Assume that $i\in\Phi$ is a non-decreasing function such that $x\mapsto x-i(x)$ is also non-decreasing.
Then, for any $T\in\mathbb{N}$
the function $x\mapsto J_T(x,i)$ is non-decreasing and continuous.
\end{lem}

\begin{pf} Note that $i$ is  Lipschitz continuous with constant less than or equal to $1.$
We proceed by induction. For $T=1$ the assertion is true by (U1). Assume that it holds for some $T\in \mathbb{N}.$
  Then,
  \begin{equation} \label{incr_1}
  J_{T+1}(x,i)=  u(x-i(x))-\frac\beta\gamma
  \ln\int_{{\mathbb R}_+}e^{-\gamma J_T(f(i(x),z),i)}\nu(dz).\end{equation}
  Now the conclusion follows as in Lemma \ref{basic} with assumption (U1). \hfill $\Box$
\end{pf}

\begin{lem} \label{covl}
Assume that $g_i$ are non-decreasing and non-negative for $i=1,2.$ Then, it follows that
\begin{eqnarray*}
\lefteqn{
-\frac1\gamma\ln\int_{{\mathbb R}_+}e^{-\gamma\left( g_1(f(y,z))+ g_2(f(y,z))\right)}\nu(dz)\le}\\&&
-\frac1\gamma\ln\int_{{\mathbb R}_+}e^{-\gamma g_1(f(y,z))}\nu(dz)
-\frac1\gamma\ln\int_{{\mathbb R}_+}e^{-\gamma g_2(f(y,z))}\nu(dz).
\end{eqnarray*}
\end{lem}

\begin{pf} The inequality follows from
 Proposition \ref{cov} in the Appendix. It suffices
to define $X:=f(y,\xi),$ where $\xi$ is a random variable of the distribution $\nu$ and put $h:=e^{-\gamma g_1},$
$g:=e^{-\gamma g_2}.$
 \hfill $\Box$
\end{pf}

For any $v\in {\cal B},$ we  define the operator $L$  as follows
\begin{equation}
\label{oper_1} Lv(x):=\sup_{y\in [0,x]} \left(u(x-y) -\frac\beta\gamma
  \ln\int_{{\mathbb R}_+}e^{-\gamma v(f(y,z))}\nu(dz)\right)
  \end{equation}
for all $x\in {\mathbb R}_+.$

 \begin{lem} \label{oper}
The operator $L$ maps ${\cal B}$ into itself and is contractive.
\end{lem}

\begin{pf} Let $v\in {\cal B}.$ First note that
by (U2) and (F2), we obtain
$$ \|Lv\|_w\le d+\alpha\beta\|v\|_w,$$
and by (U1) and Lemma \ref{basic} we have that $Lv\ge 0$ (see also (\ref{L_ogr})).
Moreover, by (U1) and Lemma \ref{basic}
the function $(x,y)\mapsto u(x-y)+\beta\widehat{v}(y)$ is continuous on $D.$ Thus,
the maximum theorem (see \cite{b}) implies that $Lv$ is continuous. Next, we observe that for $x'<x'',$
we have
$$Lv(x')\le\sup_{y\in[0,x']}\left(u(x''-y)+\beta\widehat{v}(y) \right)\le Lv(x'').$$
We now show that $Lv$ is concave (see also Sec. 2.4.4 in \cite{bra}).
Let $\lambda\in(0,1),$  $x',x''\in\mathbb{R}_+$
and $x:=\lambda x'+(1-\lambda) x''.$ By $y'\in [0,x']$ and  $y''\in [0,x'']$ we denote
the points  that attain the maximum in (\ref{oper_1}) at $x'$ and  $x'',$ respectively.
 Then, $y:=\lambda y'+(1-\lambda) y''\in [0,x].$ Hence, we get
\begin{equation} \label{oper_L}
Lv(x)\ge u(x-y)-\frac\beta\gamma
  \ln\int_{{\mathbb R}_+}e^{-\gamma v(f(y,z))}\nu(dz).
\end{equation}
Moreover, by (U1) we obtain
\begin{equation} \label{oper_u}
u(x-y)> \lambda u(x'-y')+(1-\lambda)u(x'-y').
\end{equation}
Now combining (\ref{oper_L}) with (\ref{oper_u}) and (\ref{oper_rho}) we finally obtain
$$Lv(x)> \lambda Lv(x')+(1-\lambda)Lv(x'').$$
It only remains to prove that $L$ is contractive.
Assume that $v_1,v_2\in {\cal B}.$ Then,
\begin{eqnarray*}
\lefteqn{Lv_1(x)-Lv_2(x)\le \sup_{y\in [0,x]}\left(-\frac \beta\gamma
  \ln\int_{{\mathbb R}_+}e^{-\gamma v_1(f(y,z))}\nu(dz)+\frac \beta\gamma
  \ln\int_{{\mathbb R}_+}e^{-\gamma v_2(f(y,z))}\nu(dz) \right)}\\ \nonumber
&\le& \beta\sup_{y\in [0,x]}\left(-\frac 1 \gamma
  \ln\int_{{\mathbb R}_+}e^{-\gamma \|v_1-v_2\|_w w(f(y,z))-\gamma v_2(f(y,z))}\nu(dz)+\frac 1\gamma
  \ln\int_{{\mathbb R}_+}e^{-\gamma v_2(f(y,z))}\nu(dz) \right)\\ \nonumber
&\le& \beta\sup_{y\in [0,x]} \left(- \frac1 \gamma
  \ln\int_{{\mathbb R}_+}e^{-\gamma \|v_1-v_2\|_w w(f(y,z))}\nu(dz)
   \int_{{\mathbb R}_+}e^{-\gamma v_2(f(y,z))}\nu(dz)\right.\\ \nonumber
   &&\left.+\frac 1\gamma
  \ln\int_{{\mathbb R}_+}e^{-\gamma v_2(f(y,z))}\nu(dz) \right)\\ \nonumber
&=& \beta\sup_{y\in [0,x]} -\frac1 \gamma
  \ln\int_{{\mathbb R}_+}e^{-\gamma \|v_1-v_2\|_w w(f(y,z))}\nu(dz)\\ \nonumber
&\le&   \beta\sup_{y\in [0,x]} \int_{{\mathbb R}_+} \|v_1-v_2\|_w w(f(y,z))\nu(dz)\
\mbox{ (by Jensen's inequality)} \\
&\le&   \alpha\beta\|v_1-v_2\|_w w(x) \ \ \mbox{ (by (F2))}.
\end{eqnarray*}
The second inequality follows from property (P1) and the third one from Lemma \ref{covl}
($g_1=\|v_1-v_2\|_w w,$ $g_2=v_2$) and  the fact that $w$ and $v_2$ are non-decreasing.
By changing the roles of $v_1$ with $v_2$ we obtain
$$\|Lv_1-Lv_2\|_w\le \alpha\beta\|v_1-v_2\|_w,$$
where $\alpha\beta<1.$ \hfill $\Box$
\end{pf}

\begin{pot1}  Part (a)  follows from Lemma \ref{oper} and the Banach fixed point theorem
applied to the operator $L.$
Hence, (\ref{oe}) holds. In addition, note that $V$ is strictly concave,  because of strict inequality in
(\ref{oper_u}).

Since, by Lemma \ref{basic} and (U1) the function
$$y\mapsto u(x-y)-
  \frac\beta\gamma\ln\int_{{\mathbb R}_+}e^{-\gamma V(f(y,z))}\nu(dz)$$
  is continuous and strictly concave on $[0,x]$ for $x\in\mathbb{R}_+,$
  it follows that there exists a unique point $y^*\in[0,x]$
  that realises the maximum on the right-hand side
  of (\ref{oe}).  Hence, by the maximum theorem (see \cite{b})
  there exists a unique continuous function $i^*\in\Phi$ attaining the maximum in (\ref{oe}).
In addition, strict concavity of $u$ and Lemma 3.2 in \cite{bjnjota} (see also Theorem 6.3 in  \cite{topkis})
imply that the function $i^*$
 is non-decreasing. Furthermore, observe that (\ref{oe}) can be re-formulated as follows
$$V(x)=\sup_{a\in [0,x]}\left(u(a)-
  \frac\beta\gamma\ln\int_{{\mathbb R}_+}e^{-\gamma V(f(x-a,z))}\nu(dz)\right),\quad x\in{\mathbb R}_+.$$
Assumption (U1) and strict concavity of $V$ and
 Lemma 3.2 in \cite{bjnjota} yield
   that the consumption strategy $c^*\in\Phi$
    is also non-decreasing. Clearly, $c^*(x)+i^*(x)=x$ for every $x\in{\mathbb R}_+.$
    Hence, part (b) holds true.

 Part (c). From (\ref{oe}) it follows that
 $$ V(x)\ge u(x-y) -\frac\beta\gamma\ln\int_{\mathbb{R}_+} e^{-\gamma V(f(y,z))}\nu(dz), \quad y\in[0,x] $$
 and $x\in\mathbb{R}_+.$
Let $\pi=(\pi_k)_{k\in\mathbb{N}}\in\Pi$ be any investment policy. Then, for any history $h_k\in H_k,$ $k\in\mathbb{N},$
the above display implies that
\begin{equation}
\label{L_iter}
 V(x_k)\ge L_{\pi_k}V(h_k).
 \end{equation}
 Fix any $T\in\mathbb{N}.$ Starting from (\ref{L_iter}) for $k: =T$ and applying (\ref{L_iter}) consecutively for $k=T-1,\ldots,1$
 we infer that
 $$V(x)\ge (L_{\pi_1}\circ\ldots\circ L_{\pi_T})V(x).$$
Since $V\ge 0$ and  $L_{\pi_k}$ is monotone for every $\pi_k,$ $k\in\mathbb{N},$ we obtain
\begin{equation}
\label{11}
V(x)\ge (L_{\pi_1}\circ \ldots \circ L_{\pi_T} )V(x)\ge  (L_{\pi_1}\circ \ldots \circ L_{\pi_T} ){\bf 0}(x) =J_{T}(x,\pi),
\end{equation}
for any $\pi\in\Pi$ and $x\in\mathbb{R}_+.$
Letting $T\to\infty$  in (\ref{11}), we finally have $V(x)\ge J(x,\pi)$ for any $\pi\in\Pi$ and
$x\in\mathbb{R}_+.$
Hence,
\begin{equation}
\label{it}
V(x)\ge \sup_{\pi\in\Pi}J(x,\pi)\quad x\in \mathbb{R}_+.
\end{equation}
Let $i^*\in\Phi$ be as in (\ref{oe_max}). For convenience of notation we set
$u_{i^*}(x):=u(x-i^*(x))$ and for any non-negative function $\varphi\in B_w$
$$\rho_{i^*,x}(\varphi):=- \frac1\gamma\ln\int_{{\mathbb R}_+}e^{-\gamma \varphi(f(i^*(x),z))}\nu(dz),\quad
L_{i^*}\varphi(x):=u_{i^*}(x)+\beta\rho_{i^*,x}(\varphi), \quad x\in  \mathbb{R}_+.$$
Thus,  the right hand-side of
 (\ref{oe_max}) equals
$$L_{i^*}V(x)=u_{i^*}(x)+\beta\rho_{i^*,x}(V),\qquad x\in \mathbb{R}_+. $$
By iterating the latter equality $T-1$ times we get that
\begin{equation}
\label{L_iter_i}
 V(x)= L_{i^*}^{(T)}V(x), \qquad x\in \mathbb{R}_+,
 \end{equation}
where $L_{i^*}^{(T)}$ denotes the $T$-th composition of the operator $L_{i^*}$ with itself.
Thus, (\ref{L_iter_i}), property (P1) and  Jensen's inequality together with (F2) (see also (\ref{rho_ogr}))
 yield that
\begin{eqnarray}
\label{it2}
V(x)&=& L_{i^*}^{(T-1)} (u_{i^*}
+\beta\rho_{i^*,\cdot}(V))(x) \le
 L_{i^*}^{(T-1)}
(u_{i^*}+\beta\rho_{i^*,\cdot}(\|V\|_ww))(x)\\ \nonumber
&\le&
L_{i^*}^{(T-1)}  (u_{i^*}+\alpha\beta\|V\|_w w)(x)\\ \nonumber
&=&
L_{i^*}^{(T-2)} (u_{i^*}+\beta \rho_{i^*,\cdot} (u_{i^*}+\alpha\beta\|V\|_w w))(x)
\end{eqnarray}
for $x\in \mathbb{R}_+.$
Now by putting $g_1:=u_{i^*}=J_1(\cdot,i^*),$ $g_2:=\alpha\beta\|V\|_ww$ in Lemma
 \ref{covl}, we have  that
  \begin{eqnarray}\label{it1}
  \beta\rho_{i^*,x'}(u_{i^*}
+\alpha\beta\|V\|_ww)&\le& \beta\rho_{i^*,x'}(u_{i^*})+ \beta\rho_{i^*,x'}(\alpha\beta\|V\|_ww)\\\nonumber
&\le& \beta\rho_{i^*,x'}(u_{i^*})+(\alpha\beta)^2\|V\|_ww(x'), \quad x'\in\mathbb{R}_+,
\end{eqnarray}
where the second inequality is due to Jensen's inequality and assumption (F2) (see also (\ref{rho_ogr})).
Hence, inequalities (\ref{it2}) and (\ref{it1}) combined together get
\begin{eqnarray}
\label{it3}
V(x)&\le&L_{i^*}^{(T-2)} (u_{i^*}+ \beta\rho_{i^*,\cdot}(u_{i^*})+(\alpha\beta)^2\|V\|_ww)(x)\\\nonumber&=&
L_{i^*}^{(T-3)} (u_{i^*}+ \beta\rho_{i^*,\cdot} (u_{i^*}+ \beta\rho_{i^*,\cdot}(u_{i^*})+(\alpha\beta)^2\|V\|_ww))(x).
\end{eqnarray}
We repeat the procedure. From Lemma \ref{incr}  the function
$$x\mapsto J_2(x,i^*)=u_{i^*}(x)+\beta\rho_{i^*,x}(u_{i^*})$$ is non-decreasing. Hence, making use  again of
Lemma \ref{covl} ($g_1:=J_2(\cdot,i^*),$ $g_2:=(\alpha\beta)^2\|V\|_ww$), Jensen's inequality and (F2) we get
\begin{eqnarray}
\label{it4}
\lefteqn{\beta\rho_{i^*,x'} (u_{i^*}+ \beta\rho_{i^*,\cdot}(u_{i^*})+(\alpha\beta)^2\|V\|_ww)}\\\nonumber
 &&= \beta\rho_{i^*,x'}(J_2(\cdot,i^*)+(\alpha\beta)^2\|V\|_ww)
\le \beta\rho_{i^*,x'}(J_2(\cdot,i^*))+\beta\rho_{i^*,x'}((\alpha\beta)^2\|V\|_ww) \\\nonumber
&& \le \beta\rho_{i^*,x'}(J_2(\cdot,i^*))+(\alpha\beta)^3\|V\|_ww(x'),\quad x'\in\mathbb{R}_+.
\end{eqnarray}
By combining (\ref{it3}) and (\ref{it4}) we obtain that
\begin{eqnarray*}
\label{it5}\nonumber
V(x)&\le&  L_{i^*}^{(T-3)}
 (u_{i^*}+ \beta\rho_{i^*,\cdot}(J_2(\cdot,i^*))+(\alpha\beta)^3\|V\|_ww)(x)\\
 &=& L_{i^*}^{(T-3)}(J_3(\cdot,i^*)+(\alpha\beta)^3\|V\|_ww)(x)\\
 &=& L_{i^*}^{(T-4)}  (u_{i^*}+ \beta\rho_{i^*,\cdot}(J_3(\cdot,i^*)+(\alpha\beta)^3\|V\|_ww))(x).
\end{eqnarray*}
Repeating this procedure, i.e., making use of Lemma \ref{covl}  for the functions $g_1=J_k(\cdot,i^*)$
(by Lemma \ref{incr} it  is non-decreasing)  and $g_2:=(\alpha\beta)^k\|V\|_ww$ for $k=3, \ldots, T-1$
 we finally deduce
\begin{equation}
\label{ii}
V(x)\le J_{T}(x,i^*)+(\alpha\beta)^{T}\|V\|_ww(x),\quad x\in\mathbb{R}_+.
\end{equation}
Thus, letting $T\to\infty$ in (\ref{ii}) it follows that
\begin{equation}
\label{itit}
V(x)\le J(x,i^*)\quad x\in \mathbb{R}_+.
\end{equation}
Now, (\ref{it}) and (\ref{itit}) combined together yield part (c). \hfill $\Box$
\end{pot1}

\section{The Euler Equation}

This section is devoted to establish the Euler equation.
Therefore,  we shall need additional conditions
that guarantee differentiability of functions describing the model.

\begin{enumerate}
  \item[(U3)] The function $u:\mathbb{R}_+\mapsto \mathbb{R}_+$ is continuously differentiable
  on ${\mathbb R}_{++}.$
  \item[(U4)] $u'_+(0)=\infty.$
  \item[(F3)] The  function $f(\cdot,z):\mathbb{R}_+ \mapsto \mathbb{R}_+$ is
   continuously differentiable on $\mathbb{R}_{++}.$
     \item[(F4)] $f(0,z)=0$ for all $z\ge 0.$
    \item[(F5)] There is an investment  $y>0$ such that
  $$\int_{\mathbb{R}_+} f'(y,z)\nu(dz)>0,$$
  where $f'(y,z):=\frac{\partial f(y,z)}{\partial y}.$ \footnote{Note that by Theorem 7.4 in
  \cite{slp} it follows that the function $z\mapsto f'(y,z)$ is Borel measurable.}
 \end{enumerate}

Assumption (F5) together with (F1) rules out the trivial case that $f(y,z)=0$ $\nu$-a.s. for every
$y\in\mathbb{R_+}.$

\begin{thm}\label{thm2} Assume (U1)-(U4) and (F1)-(F5). Then, we have the following.
\begin{itemize}
 \item[(a)]  For any $x\in {\mathbb R}_{++}$ the Euler equation holds
  \begin{eqnarray}
  \label{e}
  u'(c^*(x))=\beta \frac{\int_{\mathbb{R}_+} e^{-\gamma
  V(f(i^*(x),z))}u'(c^*(f(i^*(x),z)))f'(i^*(x),z)\nu(dz)}
  {\int_{{\mathbb R}_+}e^{-\gamma V(f(i^*(x),z))}\nu(dz)},
  \end{eqnarray}
  where $V$ is the function obtained in Theorem \ref{thm1}.
  \item[(b)]  The functions $x\mapsto i^*(x)$ and $x\mapsto c^*(x)$ are increasing.
\end{itemize}
\end{thm}

\begin{rmk} The Euler equation for the agent with risk sensitive preferences incorporates, in contrast to the standard expected utility case,
the value function $V.$ When $\gamma= 0,$ then equation (\ref{e}) becomes the well-known Euler equation for the model
with the expected utility, see \cite{bm} or \cite{kami}.
\end{rmk}

Let us define
\begin{equation}
\label{b}
\widehat{V}(y):=-\frac 1\gamma\ln\int_{\mathbb{R}_+} e^{-\gamma V(f(y,z))}\nu(dz).
\end{equation}
In the subsequent lemmas we shall assume that conditions (U1)-(U4) and (F1)-(F5) hold true.

 \begin{lem} \label{lem_b}
 The function $\widehat{V}$ defined in (\ref{b}) is concave, continuous and non-decreasing. Moreover,
 \begin{equation} \label{lem_bb}
\widehat{V}'_+(0)=\infty.
\end{equation}
\end{lem}

\begin{pf} The first part follows from Lemma \ref{basic}.
Take any sequence  $y_n\to 0^+$ as $n\to\infty.$
By (F4) and the fact that $u(0)=0$ we get that $V(0)=0$ and $\widehat{V}(0)=0.$
Hence,
\begin{equation}\label{lem_b0}
\frac{\widehat{V}(y_n)-\widehat{V}(0)}{y_n}= -\frac1{\gamma y_n}
  \ln\int_{{\mathbb R}_+}e^{-\gamma V(f(y_n,z))}\nu(dz)
  \ge -\frac1{\gamma}
  \ln\int_{{\mathbb R}_+}e^{-\gamma\frac{ u(f(y_n,z))}{y_n}}\nu(dz).
\end{equation}
From the chain rule,  we have for any $z\in\mathbb{R}_+$ that
\begin{equation}\label{lem_b1}
\frac{ u(f(y_n,z))}{y_n}=\frac{ u(f(y_n,z))-u(0)}{f(y_n,z)-f(0,z)}\frac{f(y_n,z)-f(0,z)}{y_n}.
\end{equation}
Letting $n\to\infty$ in (\ref{lem_b1}), we obtain that
\begin{equation}\label{lem_b2}
\lim_{n\to\infty}\frac{ u(f(y_n,z))}{y_n}= u'_+(f(0,z))f'_+(0,z).
\end{equation}
Note that the convergence in (\ref{lem_b2}) is monotonic, since $u$ and $f(\cdot,z)$
are concave and $f(\cdot,z)$ is non-decreasing  for $z\in\mathbb{R}_+$.
By the monotone convergence theorem, (\ref{lem_b0})  and (\ref{lem_b2}) we finally get
$$\widehat{V}'_+(0)\ge -\frac1{\gamma}
  \ln\int_{{\mathbb R}_+}e^{-\gamma u'_+(0)f'_+(0,z)}\nu(dz).$$
Assumption (F5) together with (F1) yield that $f'_+(0,z)>0.$ Thus, by (U4) the assertion follows. \hfill $\Box$
\end{pf}

 \begin{lem} \label{lem_kami}
Let $i^*$ be defined in (\ref{oe_max}). Then, $i^*(x)\in (0,x)$ for any $x\in\mathbb{R}_{++}.$
\end{lem}

\begin{pf} The assertion that $i^*(x)>0$ follows from assumption (U4),
whereas  (\ref{lem_bb}) yields $i^*(x)<x.$ The reader is referred to Lemma 5 in \cite {kami}. \hfill $\Box$
\end{pf}

 \begin{lem} \label{lem_env}
The function $V$  is continuously differentiable on $\mathbb{R}_{++}$ and
$V'(x)=u'(c^*(x)),$ for $x\in\mathbb{R}_{++}$.
\end{lem}

\begin{pf} The way of showing the equality proceeds along the same lines as the proofs of
Proposition 12.1.18 and Corollary 12.1.19 in \cite{st}.  \hfill $\Box$
\end{pf}

\begin{pot2} First we show part (a). In view of Lemma \ref{lem_env} and (\ref{oe}),
it suffices to show that $\widehat{V}$ is differentiable
on $\mathbb{R}_{++}$ and $\beta \widehat{V}'(y)$ at $y=i^*(x)$ equals to the right-hand side of (\ref{e}).
Since $\widehat{V}$ is concave by Lemma \ref{lem_b},
we know that the right-hand side and the left-hand side derivatives exist.
Let $y\in\mathbb{R}_{++}$ be arbitrary and $h>0.$ Set
$$F(y,z):=\frac {e^{-\gamma V(f(y,z))}}{\int_{{\mathbb R}_+}e^{-\gamma V(f(y,z))}\nu(dz)}>0$$
and note that
$$\int_{{\mathbb R}_+} F(y,z)\nu(dz)=1\quad \mbox{ for any }\ y\in \mathbb{R}_{++}.$$
Then, by Lemma \ref{lem_env} and (F4)
\begin{eqnarray*}
\frac{\widehat{V}(y+h)-\widehat{V}(y)}{h}&=&
-\frac 1{\gamma h}\ln \frac {\int_{{\mathbb R}_+}e^{-\gamma V(f(y+h,z))}\nu(dz)}
{\int_{{\mathbb R}_+}e^{-\gamma V(f(y,z))}}\\
&=& -\frac 1{\gamma h}\ln \int_{{\mathbb R}_+}e^{-\gamma (V(f(y+h,z))-V(f(y,z)))}F(y,z)\nu(dz) \\
&\le& \int_{{\mathbb R}_+}\frac{V(f(y+h,z))-V(f(y,z))}hF(y,z)\nu(dz)\\
&=& \int_{{\mathbb R}_+}\frac{V(f(y+h,z))-V(f(y,z))}{f(y+h,z)-f(y,z)}\frac{f(y+h,z)-f(y,z)}hF(y,z)\nu(dz)\\
&\le&  \int_{{\mathbb R}_+} V'(f(y,z))f'(y,z) F(y,z)\nu(dz)=: G(y),
\end{eqnarray*}
 where the first inequality is due to Jensen's inequality and the second one follows from
 the fact that $V$ and $f(\cdot,z)$ for $z\in\mathbb{R}_+$ are concave and non-decreasing.
Thus, for $y\in\mathbb{R}_{++}$
\begin{equation}
\label{t21}
\widehat{V}'_+(y)\le G(y).
\end{equation}
Let us now consider the left-hand side derivative of $\widehat{V},$ i.e.,
\begin{eqnarray*}
\frac{\widehat{V}(y-h)-\widehat{V}(y)}{-h}&=&
\frac 1{\gamma h}\ln \frac {\int_{{\mathbb R}_+}e^{-\gamma V(f(y-h,z))}\nu(dz)}
{\int_{{\mathbb R}_+}e^{-\gamma V(f(y,z))}}\\
&\ge& \int_{{\mathbb R}_+}\frac{V(f(y-h,z))-V(f(y,z))}{-h}F(y,z)\nu(dz)\\
&=&
\int_{{\mathbb R}_+}\frac{V(f(y-h,z))-V(f(y,z))}{f(y-h,z)-f(y,z)}\frac{f(y-h,z)-f(y,z)}{-h}F(y,z)\nu(dz)\\
&\ge&  \int_{{\mathbb R}_+} V'(f(y,z))f'(y,z) F(y,z)\nu(dz)= G(y).
\end{eqnarray*}
Hence, for $y\in\mathbb{R}_{++}$
\begin{equation}
\label{t22}
\widehat{V}'_-(y)\ge G(y).
\end{equation}
Observe that $G$ is continuous. This is due to  Lemma \ref{lem_env}, (F1), (F4)
and the dominated convergence theorem.
Since $\widehat{V}$ is concave and (\ref{t21}) and (\ref{t22}) hold, then for $h>0$ we obtain that
\begin{equation}\label{g} G(y+h)\le \widehat{V}'_-(y+h)\le \widehat{V}'_+(y)
\le G(y)\le \widehat{V}'_-(y)\le \widehat{V}'_+(y-h)\le G(y-h).
\end{equation}
Now letting $h\to 0^+$ in (\ref{g}), it follows that $\widehat{V}$ is
continuously differentiable on $\mathbb{R}_{++}.$

In order to prove (b) suppose that $x'< x''.$ If $x'=0,$ then by Lemma \ref{lem_kami} we have
$i^*(x')=0<i^*(x'')$ and $c^*(x')=0<c^*(x'')=x''-i^*(x'').$
Hence, let  $x'>0$ and $i^*(x')=i^*(x'').$ Then, by Euler equation (\ref{e}), we obtain
$u'(x'-i^*(x'))=u'(x''-i^*(x')).$ But the equality cannot hold,
since  $u$ is strictly concave. Similarly, if
$c^*(x')=c^*(x''),$ then by Lemma \ref{lem_env} we must have
$V'(x')=u'(c^*(x'))=u'(c^*(x''))=V'(x'').$
However, this equality contradicts the strict concavity of $V.$ \hfill $\Box$
\end{pot2}

\section{Stationary Distributions}

In this section, we shall consider dynamics of the growth model when the agent follows the optimal  policy
$i^*\in \Phi.$ Our aim is to provide a set of assumptions under which the system is globally stable and the resulting
stationary distribution  is non-trivial in the sense that it is not concentrated on zero. We wish to follow the approach
studied in \cite{ns_jet} and further developed by \cite{kami} and \cite{st}.
It combines the Euler equation (see (\ref{e})) with the Foster-Lyapunov theory of Markov chains.

More precisely, we deal with the process
\begin{equation}\label{optimal}
x_{t+1}=f(i^*(x_t),\xi_{t}),\qquad (\xi_t)_{t\in\mathbb{N}}\mbox{ is i.i.d. sequence, where } \xi_t\sim \nu\in \Pr(\mathbb{R}_+)
\end{equation}
and $f: \mathbb{R}_+\times \mathbb{R}_+\mapsto \mathbb{R}_+$ is continuous. Clearly, $ (x_t)_{t\in\mathbb{N}}$ is a Markov process.
Assuming that $f(y,z)>0$ for every $y\in\mathbb{R}_{++}$
and utilising the fact that $i^*(x)\in(0,x)$ for $x>0,$
we may confine ourselves to the study of the income process on $\mathbb{R}_{++}$ (see p. 303 in \cite{st}).
We show the existence of at least one stationary non-trivial distribution. According to Proposition  \ref{pr2} in the Appendix,
we have to find a function  $W: \mathbb{R}_{++}\mapsto \mathbb{R}_{+}$ satisfying properties (a) and (b).
Since we wish to avoid repeating all the details contained in the aforementioned papers, we focus only on an element in the proof that makes
 use of the Euler equation. This is because, the Euler equation in our framework  is different than the one obtained for the  expected utility case.
Therefore, we formulate only crucial
conditions in the most simple way. The reader is referred to  \cite{kami} and  \cite{ st}, where
 a detailed discussion and more general conditions are provided  implying
the ones given below.

\begin{enumerate}
  \item[(D1)]  It is satisfied that
 $$\lim_{y\to 0^+}\int_{{\mathbb R}_{+}} \frac1{\beta f'(y,z)}\nu(dz)<1.$$
  \item[(D2)] There exist $\lambda_2\in(0,1)$ and $\kappa_2>0$ such that
$$\int_{\mathbb{R}_{+}} f(y,z)\nu(dz)\le \lambda_2 y+\kappa_2,\quad y\in\mathbb{R}_+.$$
 \end{enumerate}

Here, we would like to mention that Assumption (D1) prevents probability mass from escaping to infinity, whereas the role of  (D2)
is to prevent probability mass from escaping to zero.

 \begin{lem} \label{lem_distr}
Assume (D1). Then, for $W_1(x):=\sqrt{u'(c^*(x))e^{-\gamma V(x)}},$ $x\in\mathbb{R}_{++},$ there exist
$\lambda_1\in(0,1)$ and $\kappa_1>0$ such that
$$\int_{\mathbb{R}_{+}} W_1(f(i^*(x),z))\nu(dz)\le \lambda_1 W_1(x)+\kappa_1,\quad x\in\mathbb{R}_{++}.$$
\end{lem}

\begin{pf} First note that by the Cauchy-Schwarz inequality, it follows that
\begin{eqnarray}
\label{cs}\nonumber
\lefteqn{
\int_{\mathbb{R}_{+}} W_1(f(i^*(x),z))\nu(dz)=}\\ \nonumber
&&\int_{\mathbb{R}_{+}} \left[u'(c^*(f(i^*(x),z))) e^{-\gamma
  V(f(i^*(x),z))} \frac{\beta f'(i^*(x),z)}{\beta f'(i^*(x),z)}\frac { \int_{ \mathbb{R}_+}e^{-\gamma V(f(i^*(x),z))}\nu(dz)}
  { \int_{ \mathbb{R}_+}e^{-\gamma V(f(i^*(x),z))} \nu(dz)}\right]^{\frac12}\nu(dz)\\ \nonumber
&\le& \left[\int_{\mathbb{R}_{+}} u'(c^*(f(i^*(x),z))) e^{-\gamma
  V(f(i^*(x),z))} \frac{\beta f'(i^*(x),z)} { \int_{ \mathbb{R}_+}e^{-\gamma V(f(i^*(x),z))}\nu(dz)}\nu(dz)\right]^{\frac12}\times\\
 &&\times \left[  \int_{ \mathbb{R}_+} \frac
  { \int_{ \mathbb{R}_+}e^{-\gamma V(f(i^*(x),z))} \nu(dz)}
  {\beta f'(i^*(x),z)}\nu(dz)\right]^{\frac12},\qquad x\in\mathbb{R}_{++}.
\end{eqnarray}
 Furthermore,  applying   (\ref{e}) to (\ref{cs}) we obtain that
\begin{eqnarray}
\label{cs1}\nonumber
\int_{\mathbb{R}_{+}} W_1(f(i^*(x),z))\nu(dz)
  &\le& \sqrt{u'(c^*(x))}\left[
\int_{\mathbb{R}_{+}}\frac{ \nu(dz)} {\beta f'(i^*(x),z)}
 \int_{ \mathbb{R}_+}e^{-\gamma V(f(i^*(x),z))} \nu(dz)\right]^{\frac12}\\
   &\le& \sqrt{u'(c^*(x))}\left[
\int_{\mathbb{R}_{+}}\frac{ \nu(dz)} {\beta f'(x,z)} \right]^{\frac12},
\end{eqnarray}
where the second inequality is due to the fact that $V\ge 0$ and (F1) ($f'(\cdot,z)$ is non-increasing for $z\in\mathbb{R}_+$).
From assumption (D1), it follows that there exists $\delta>0$ such that
$$\lambda_1:= e^{\gamma/2V(\delta)}\left[\int_{\mathbb{R}_{+}}\frac{ \nu(dz)} {\beta f'(\delta,z)}\right]^{\frac12}<1.$$
Then, for $x\in(0,\delta)$  we have that $e^{-\gamma/2V(x)}e^{\gamma/2V(\delta)}\ge 1,$ and consequently, by (\ref{cs1})
\begin{equation}
\label{81}
\int_{\mathbb{R}_{+}} W_1(f(i^*(x),z))\nu(dz)
  \le \sqrt{u'(c^*(x))e^{-\gamma V(x)}} e^{\gamma/2V(\delta)}\left[
\int_{\mathbb{R}_{+}}\frac{ \nu(dz)} {\beta f'(\delta,z)} \right]^{\frac12}\le \lambda_1 W_1(x).
\end{equation}
For $x\ge \delta$ we have
\begin{equation}
\label{82}
\int_{\mathbb{R}_{+}} W_1(f(i^*(x),z))\nu(dz) \le \int_{\mathbb{R}_{+}}\left[u'(c^*(f(i^*(\delta),z))) e^{-\gamma
  V(f(i^*(\delta),z))}\right]^{\frac 12}\nu(dz)=:\kappa_1.
\end{equation}
Inequalities (\ref{81}) and (\ref{82}) combined together yield the conclusion.
  \hfill $\Box$
\end{pf}

From Lemma \ref{lem_distr} and (D2) we deduce that the function
$$W(x)=W_1(x)+x, \qquad x\in\mathbb{R}_{++}$$
satisfies point (b) in Proposition \ref{pr2} with $\lambda:=\max\{\lambda_1,\lambda_2\}$  and $\kappa:=\kappa_1+\kappa_2.$
Clearly,  from (U4) it follows that condition (a) is also satisfied. Hence, there exists a non-trivial distribution for the Markov process
in (\ref{optimal}).

The  problem of the  uniqueness of a stationary distribution, i.e., global stability,  has been recently
 studied  in \cite{kami},  \cite{ns_jet}  and \cite{st}.
Therefore, we do not give here all these assumptions and refer the reader to the above-mentioned works. Further results on invariant
distributions  are also widely reported   in \cite{mt}, \cite{bm_book} and in the recent  papers of \cite{ks}, \cite{z}.

\begin{rmk}
If the income process evolves on the compact space $[0,\bar{s}],$
then the assumptions (U2) and (F2) are satisfied with $w\equiv 1.$
Consequently, the results in Sections 3 and 4 are satisfied, in particular, the optimal investment policy is non-deacreasing.
In this case, the existence of a non-trivial invariant  distribution follows from the Krylov-Bogolubov theorem, see for instance,
Theorem 11.2.5 in \cite{st}.
\end{rmk}

\section{Examples}

Below we provide two examples of utility and production functions that meet our assumptions used in Sections 3-4.

 \begin{ex} \label{ex_multi} {\it A model with multiplicative shocks.}
 Assume that the process
 evolves according to difference equation
 $$x_{t+1}=y_t^\theta\xi_t, \quad t\in\mathbb{N},$$
where $\theta\in(0,1).$ Suppose that $\bar{z}:=\int_{\mathbb{R}_+}z\nu(dz)$ is finite and let
$u(a)=a^\sigma$ with $\sigma\in(0,1).$ Clearly, (U1) and (F1) are satisfied. Assumption (U2) holds
for $w(x)=(r+x)^\sigma,$ where $r\ge 1$ is a constant sufficiently large so that
$$\left( 1+\frac{\bar{z}^{\frac 1{1-\theta}}}r \right)^\sigma\beta<1.$$
Then, calculations on p. 263 in \cite{jn} show that (F2) is also satisfied with
$\alpha:=\left( 1+\frac{\bar{z}^{ 1/1-\theta}}r \right)^\sigma.$
We also note that (U3)-(U4) and (F3)-(F5)  hold true.

For this model,  conditions (D1) and (D2) are met as well.  Clearly, the  finiteness of  $\int_{\mathbb{R}_+}1/z\nu(dz)$ implies (D1).
Define now
$$\kappa_1:=\max\{\bar{z}, \bar{z}^{\frac 1{1-\theta}}(1-\theta)\},\quad \lambda_1:=\theta.$$
Then, for $y\le 1$ we have
$$\int_{\mathbb{R}_+}y^\theta z\nu(dz)=y^\theta \bar{z}\le\bar{z}\le \theta y +\kappa_1.$$
Set  $l(y):= \theta y-\bar{z}y^\theta+\kappa_1$ for $y>0.$ This function attains minimum at
$$y_{\min}=\bar{z}y^{\frac1{1-\theta}}\quad\mbox{and}\quad l(y_{\min})\ge 0.$$
Hence,
$$\int_{\mathbb{R}_+} f(y,z)\nu(dz)=\bar{z}y^\theta\le  \lambda_1 y+\kappa_1$$
for all $y\in\mathbb{R}_+.$
 \end{ex}

 \begin{ex} \label{ex_additive} {\it A model with additive shocks.}
Let the evolution of the process be described by the equation
 $$x_{t+1}=
 \left\{\begin{array}{l@{\quad \quad}l}
 \eta y_t+\xi_t,& \mbox{ if } y_t>0,\\
0, & \mbox{ if } y_t=0,
 \end{array}\right.
  \quad t\in\mathbb{N},$$
 where $\eta>0$ denotes a constant rate of growth. Obviously, (F1), (F3)-(F5) holds.
Let the utility $u$ be defined as in Example \ref{ex_multi}. Then, (U2) holds for $w(x)=(x+r)^\sigma$
with any $r\ge 1.$ Now, we prove assumption (F2). Namely, by the Jensen inequality it follows that
$$\sup_{y\in[0,x]}\int_{\mathbb{R}_+} (\eta y+z)\nu(dz)\le (\eta x+\bar{z}+r)^\sigma.$$
Thus,
$$\sup_{x\in \mathbb{R}_+}\frac{ (\eta x+\bar{z}+r)^\sigma}{w(x)}=(s(x))^\sigma,\quad\mbox{
where}\quad s(x):=\frac{\eta x+\bar{z}+r}{x+r}, \quad x\in \mathbb{R}_+.$$
We have
$$\lim_{x\to 0^+} s(x)= 1+\frac{\bar{z}}r,\quad\quad \lim_{x\to +\infty} s(x)= \eta.$$
If $\eta>1,$ then from the calculations in \cite{jn} on p. 264, it follows that for $r>\max\{1, \frac{\bar{z}}{\eta-1}\}$ condition
(F2) holds for all $\beta\in (0,1)$ for which $\beta\eta^\sigma<1$ (here $\alpha:=\eta^\sigma$).
If, on the other hand, $\eta\le 1,$ then (F2) is satisfied for each $\beta\in (0,1).$ Namely,
for the given discount factor it is enough to take sufficiently large $r\ge 1$ such that
$(1+\frac{\bar{z}}{r})^\sigma\beta<1.$ Obviously,
$\alpha:= (1+\frac{\bar{z}}{r})^\sigma.$

In this case, assumption (D1) is not met. However, the existence of an invariant distribution  can be proved under some extra
requirements with the help of other techniques, see  for instance, p. 207 and p. 259  in
\cite{st}  or \cite{mt}.
 \end{ex}

\section{Appendix}

The following result can be found in \cite{dgl} (Theorem A.19).

\begin{pr}
\label{cov}
Let $X$ be a real-valued random
variable defined on $(\Omega,{\cal F}, P)$ and let $h$ and $g$ be  non-increasing
real-valued functions. Then,
$$E\{h(X)g(X)\} \ge E\{h(X)\}E\{g(X)\},$$
provided that all expectations exist and are finite.
\end{pr}

For the proof of the next result the reader is referred  to  \cite{kami} (Lemma 3.1)
or  to  \cite{st} ( Corollary 11.2.10).

\begin{pr}
\label{pr2}
Consider the Markov process defined in (\ref{optimal}). Suppose that there exists a function
$W: \mathbb{R}_{++}\mapsto \mathbb{R}_{+}$ such that
$$\mbox{(a)}\quad
\lim_{x\to\infty} W(x)=\infty\quad\mbox{and}\quad  \lim_{x\to 0^+} W(x)=\infty$$
$$\mbox{(b)}\quad\exists \kappa>0,\;\exists \lambda\in[0,1), \forall x\in \mathbb{R_{++}}\quad
\int_{\mathbb{R}_{+}} W(f(i^*(x),z))\nu(dz)\le \lambda W(x)+\kappa.$$
Then, there exists at least one non-trivial stationary distribution on $\mathbb{R}_+.$
\end{pr}

\bibliographystyle{abbrv}

\end{document}